# An improved multicomponent pseudopotential lattice Boltzmann method for immiscible fluid displacement in porous media

M. Sedahmed[*1], R. C. V. Coelho[2,3], H. A. Warda[1]

*(1) Mechanical Engineering Department, Alexandria University, Alexandria 21544, Egypt.*
*(2) Centro de Física Teórica e Computacional, Faculdade de Ciências, Universidade de Lisboa, 1749-016 Lisboa, Portugal.*
*(3) Departamento de Física, Faculdade de Ciências, Universidade de Lisboa, 1749-016 Lisboa, Portugal.*
*[*]Corresponding author email: mahmoud.sedahmed@alexu.edu.eg*

## ABSTRACT

Immiscible fluid displacement in porous media occurs in several natural and industrial processes. For example, during petroleum extraction from porous rock reservoirs, water is used to displace oil. In this paper, we investigate primary drainage and imbibition in a heterogeneous porous medium using an improved numerical model based on the multicomponent pseudopotential lattice Boltzmann method. We apply recent developments from the literature and develop new pressure boundary conditions. We show that the proposed model is able to simulate realistic viscosity ratios, and it allows independent tuning of surface tension from viscosity. Moreover, the model suppresses a non-physical behavior of previous schemes, in which trapped fluid volumes significantly change with time. Furthermore, we show that the developed model correctly captures the underlying physical phenomena of fluid displacements. We simulate oil-water flows and verify that the measured values of irreducible water and residual oil saturations are realistic. Finally, we vary the wetting conditions of the porous medium to represent different wettability states. For the different scenarios, we show that the simulations are in good agreement with experimental results.

## KEYWORDS

Multicomponent fluid flow,
Porous media,
Wetting boundary conditions,
Pressure boundary conditions,
Trapped fluid volumes,
Fluid displacement processes,

## 1. INTRODUCTION

The study of immiscible fluid displacement processes in porous media is essential for many natural and industrial systems. One prominent example is underground petroleum reservoirs where oil and water displacement occur. For instance, oil displaces water during oil migration to the reservoir rock, and water displaces oil during water injection to increase reservoir productivity [1]. These displacement processes were described on the macroscopic scale (meters to kilometers) using the well-known Darcy equation for years. However, much underlying physics occurs on the pore scale (micrometers to millimeters) [2]. Moreover, the characterization of the different displacement processes relied on laboratory experiments that mimic reservoir conditions. For example, Special Core Analysis (SCAL) is a group of experiments that quantify vital fluid and rock properties used to describe petroleum flow in reservoirs [3]. Many relationships are established using SCAL, such as capillary pressure-saturation curves and relative permeability curves. Despite being the standard method of quantifying such relationships, SCAL experiments are expensive and time-consuming. Such experiments can be replicated numerically with recent advances in image processing, computational power, and numerical models [2].

Different approaches were used for simulating the fluids displacement process in porous media [4]. An increasingly popular approach for computational fluid dynamics (CFD) is the lattice Boltzmann method (LBM) [5] [6] [7] [8] [9] [10] [11] [12] [13] [14] [15]. LBM is based on the kinetic theory of gases allowing it to capture physical phenomena on the microscale and translate them into macroscopic parameters. In the past three decades, different LBM models were developed for simulating multicomponent fluids flow, with the most popular being the color gradient model [16], the free energy model [17], and the pseudopotential model [18]. In this work, we use the pseudopotential model, also known as the Shan-Chen (SC) model. There are mainly two versions of the SC model: single component multiphase (SCMP) and multicomponent multiphase (MCMP) [13]. A multicomponent (MC) pseudopotential model could be considered a special case of the MCMP model where phase change is not considered for the fluid components.

Several works in the literature used the MC model to simulate immiscible fluid displacement in porous media due to its simplicity. Since the early days of the MCMP model, Chen and Martys [19] used it to simulate drainage and imbibition processes in a three-dimensional (3D) porous medium reconstructed from a microtomography image Fontainebleau sandstone. Pan et al. [20] simulated fluids flow a porous medium comprised of a synthetic packing, and they achieved good agreement between numerical and experimental results. Schaap et al. [21] simulated fluids displacement processes in a glass bead porous medium obtained from computed tomography and quantitatively compared the numerical results with experimental results. Porter et al. [22] investigated hysteresis in the relationship between capillary pressure $P_c$ and wetting phase saturation $S_w$ for drainage and imbibition processes in a similar porous medium. Zhao et al. [23] simulated the flow of two immiscible fluids in a two-dimensional (2D) porous medium. They analyzed the effects of capillary number (Ca), viscosity ratio (M), and wettability on the relative permeability curves. Warda et al. [24] simulated the primary drainage and imbibition displacement processes in a 2D generated heterogeneous porous medium. They showed that the MC model captured the capillary pressure bump phenomenon due to heterogeneity of the porous medium. Fager et al. [25] presented a digital SCAL workflow based on LBM simulations to directly simulate an enhanced oil recovery (EOR) technique named water alternating gas (WAG) on a digitized rock geometry. Nemer et al. [26] investigated the relative importance of the wettability-altered fraction, the degree of wettability alteration, the accurate contact angle assignment, and their effects on relative permeability and fluid configurations using the MCMP model.

Notwithstanding being the most widely model in the literature, the original version of the MCMP suffered from several drawbacks; for instance, it had a limited density ratio (~1), a limited viscosity ratio (~10), high spurious currents, and tuning of surface tension dependent of other quantities [11]. Moreover, it suffered from a non-physical change in the volumes of trapped fluids, as shown in Ref. [27]. Such changes negatively impacted the simulations of primary drainage and imbibition processes in porous media [28]. For the past 15 years, several enhancements have been proposed for the SC model to overcome its drawbacks. However, more studies were dedicated to the SCMP model, while the MCMP model did not receive the same attention despite its importance in industrial applications [11]. Notably, most model enhancements served the collision model [11] [12]. Recently developed models relied on combining several enhancements from the literature to represent an improved model [29] [30] [31].

Although the collision model is the workhorse of the LBM algorithm, boundary conditions play a vital role in the stability and accuracy of the model. In LBM, macroscopic boundary conditions (e.g., Dirichlet boundary conditions) are not directly implemented as in conventional CFD models where their values are directly used in the governing equations. A conversion technique is needed to construct a relationship between the macroscopic boundary conditions and the LBM's basic unit, the particle distribution function [9]. Zou and He [32] showed how such a conversion could be made, but their study was concerned with single-phase fluid flow only. Multicomponent fluids flow would need additional treatments to accommodate the flow of different components and achieve stable results, which is a challenging task. For instance, Hou et al. [33] proposed a new multiphase open boundary condition to enable fluid droplets to naturally pass the outlet boundary without deformations. As we will show in the present study, this treatment for boundary conditions did not provide adequate results for simulations of immiscible fluid displacement processes, mainly where there are volumes of trapped fluids.

Another boundary condition that has a significant role in the simulations of fluids flow in porous media is the wetting boundary condition since fluids flow behavior is greatly affected by the wettability state of the solid walls. For that, we implement a recently developed model known as the improved virtual solid density (IVSD) model [34]. That scheme was shown to be more accurate than the previous ones, especially in complex geometries such as porous media.

In this work, we introduce an improved MC pseudopotential model and show that it can alleviate many of the mentioned drawbacks of the previous schemes. The introduced model is based on a recently developed modified version of the explicit forcing (EF) scheme [35] [36]. The modified EF scheme was shown to achieve high viscosity ratios (~250) when combined with single relaxation time (SRT) approximation of the collision operator, while it could achieve even higher viscosity ratios (~1000) with multiple-relaxation time (MRT) [35]. Since we are concerned with multicomponent fluid flow where density difference could be neglected, density ratio was kept at unity, and no additional treatments were included to achieve higher density ratios. The improved collision model is combined with the recently developed IVSD model for wetting boundary conditions [34] and special treatment for fluid interaction forces at the inlet and outlet sides of the simulation domain [37]. Furthermore, we introduce a new pressure boundary condition at the inlet and outlet sides of the simulation domain.

This paper is organized as follows. In section 2, we present the developed numerical model and boundary conditions. Section 3 provides extensive numerical tests for the model using some widely used benchmarks from the literature. In section 4, we show numerical results and interpretations for the simulations of fluid displacement processes over an obstacle and in a heterogeneous porous medium. Moreover, we qualitatively compare the results with experimental observations and well-known physical phenomena in petroleum reservoirs. Finally, we provide a concise summary in section 5.

## 2. METHODOLOGY

### 2.1. Multicomponent pseudopotential lattice Boltzmann method

In this work, we employ the MC pseudopotential lattice Boltzmann method with the modified explicit forcing (EF) scheme and the SRT approximation of the collision operator [35]. The discretized Lattice Boltzmann Equation (LBE) for this model is written as

$$f_i^{(\sigma)}(\pmb{x} + \pmb{e}_i \, \Delta t, t + \Delta t) - f_i^{(\sigma)}(\pmb{x}, t)$$
$$= -\frac{1}{\tau_s} \left( f_i^{(\sigma)}(\pmb{x}, t) - f_i^{(\sigma), eq}(\pmb{x}, t) \right) + \Delta t \left( 1 - \frac{1}{2\, \tau_s} \right) S_i^{(\sigma), F}(\pmb{x}, t), \quad (1)$$

where, $f_i^{(\sigma)}$ is the particle distribution function at position $(\pmb{x})$ and time $(t)$, $\sigma$ is the fluid component number ($\sigma = 1, \dots n$). We study only two fluid components, i.e., $n = 2$. $i$ is the lattice direction that belongs to the selected lattice arrangement; Here we use the $D_2Q_9$ lattice arrangement for 2D simulations ($i = 0, 1, \dots 8$). $f_i^{(\sigma), eq}$ is the equilibrium distribution function, and it reads as

$$f_i^{(\sigma), eq}(\pmb{x}) = w_i \rho^{(\sigma)}(\pmb{x}) \left[ 1 + \frac{\pmb{e}_i \cdot \pmb{u}^{eq}(\pmb{x})}{c_s^2} + \frac{\left( \pmb{e}_i \cdot \pmb{u}^{eq}(\pmb{x}) \right)^2}{2\, c_s^4} - \frac{\pmb{u}^{eq}(\pmb{x}) \cdot \pmb{u}^{eq}(\pmb{x})}{2\, c_s^2} \right], \quad (2)$$

where, $c_s$ is the lattice sound speed, and it is defined as $c_s = \frac{\Delta x}{\sqrt{3}\,\Delta t}$. $\Delta x$ and $\Delta t$ are commonly chosen as unity in the LBM. $w_i$ and $\pmb{e}_i$ are the lattice weights and the discrete velocity vector in the $i^{th}$ direction, respectively. In this work, we use $D_2Q_9$ lattice arrangement, hence $w_i$ and $\pmb{e}_i$ are given by

$$\pmb{e}_i = \begin{bmatrix} 0 & 1 & 0 & -1 & 0 & 1 & -1 & -1 & 1 \\ 0 & 0 & 1 & 0 & -1 & 1 & 1 & -1 & -1 \end{bmatrix},$$

$$w_i = \left[\frac{4}{9} \quad \frac{1}{9} \quad \frac{1}{9} \quad \frac{1}{9} \quad \frac{1}{9} \quad \frac{1}{36} \quad \frac{1}{36} \quad \frac{1}{36} \quad \frac{1}{36}\right].$$

The macroscopic fluid component density, velocity, and mixture (total) density are given by

$$\rho^{(\sigma)}(\boldsymbol{x}) = \sum_i f_i^{(\sigma)}(\boldsymbol{x}), \tag{3}$$

$$\boldsymbol{u}^{(\sigma)}(\boldsymbol{x}) = \frac{1}{\rho^{(\sigma)}(\boldsymbol{x})} \sum_i \boldsymbol{e}_i f_i^{(\sigma)}(\boldsymbol{x}), \tag{4}$$

$$\rho(\boldsymbol{x}) = \sum_\sigma \rho^{(\sigma)}(\boldsymbol{x}). \tag{5}$$

The system relaxation time $\tau_s$ is defined as

$$\tau_s(\boldsymbol{x}) = \frac{\sum_\sigma \rho^{(\sigma)}(\boldsymbol{x}) \, \nu^{(\sigma)}}{\rho(\boldsymbol{x}) \, c_s^2 \, \Delta t} + \frac{1}{2}, \tag{6}$$

where, $\nu^{(\sigma)}$ is the kinematic viscosity of the $\sigma^{th}$ fluid component, and it is determined by

$$\nu^{(\sigma)} = c_s^2 \left(\tau^{(\sigma)} + \frac{1}{2}\right) \Delta t, \tag{7}$$

where, $\tau^{(\sigma)}$ is the relaxation time of the $\sigma^{th}$ fluid component. The system relaxation time $\tau_s$ is smooth at the interface between the two fluids in order to make the collision model more stable.

The equilibrium velocity $\boldsymbol{u}^{eq}$ reads

$$\boldsymbol{u}^{eq}(\boldsymbol{x}) = \frac{\sum_\sigma \frac{\rho^{(\sigma)}(\boldsymbol{x}) \, \boldsymbol{u}^{(\sigma)}(\boldsymbol{x})}{\tau_s(\boldsymbol{x})}}{\sum_\sigma \frac{\rho^{(\sigma)}(\boldsymbol{x})}{\tau_s(\boldsymbol{x})}}. \tag{8}$$

Also, the common velocity of the two fluid components is defined as

$$\boldsymbol{u}(\boldsymbol{x}) = \boldsymbol{u}^{eq}(\boldsymbol{x}). \tag{9}$$

The forcing term in the LBE equation $S_i^{(\sigma),F}$ is defined as follows:

$$S_i^{(\sigma),F}(\boldsymbol{x}) = w_i \left(\frac{\boldsymbol{e}_i \cdot \boldsymbol{F}^{(\sigma),tot}(\boldsymbol{x})}{c_s^2}\right.$$
$$\left. + \frac{(\boldsymbol{e}_i \boldsymbol{e}_i - c_s^2 I) : \left(\boldsymbol{u}^{eq}(\boldsymbol{x}) \boldsymbol{F}^{(\sigma),tot}(\boldsymbol{x}) + \boldsymbol{F}^{(\sigma),tot}(\boldsymbol{x}) \boldsymbol{u}^{eq}(\boldsymbol{x})\right)}{2 \, c_s^4}\right), \tag{10}$$

which can be rewritten as

$$S_i^{(\sigma),F}(\boldsymbol{x}) = w_i \left(\frac{\boldsymbol{e}_i - \boldsymbol{u}^{eq}(\boldsymbol{x})}{c_s^2} + \frac{\boldsymbol{e}_i \cdot \boldsymbol{u}^{eq}(\boldsymbol{x})}{c_s^4} \boldsymbol{e}_i\right) \cdot \boldsymbol{F}^{(\sigma),tot}(\boldsymbol{x}), \tag{11}$$

where, $\boldsymbol{F}^{(\sigma),tot}$ is the total force exerted on the $\sigma^{th}$ fluid component, and it is defined for the pseudopotential model by,

$$\boldsymbol{F}^{(\sigma),tot} = \boldsymbol{F}^{(\sigma),f-f} + \boldsymbol{F}^{(\sigma),f-s} + \boldsymbol{F}^{(\sigma),b}, \tag{12}$$

where, $\boldsymbol{F}^{(\sigma),f-f}$ is the fluid-fluid interaction force of the $\sigma^{th}$ fluid component and it is determined by,

$$\boldsymbol{F}^{(\sigma),f-f}(\boldsymbol{x}) = -G_{coh}^{(\sigma\bar{\sigma})} \psi^{(\sigma)}(\boldsymbol{x}) \sum_i w_i \, \psi^{(\bar{\sigma})}(\boldsymbol{x} + \boldsymbol{e}_i) \boldsymbol{e}_i, \tag{13}$$

where, $\psi^{(\sigma)}$ is the pseudopotential of the $\sigma^{th}$ fluid component. Here we define it as $\psi^{(\sigma)} = \rho^{(\sigma)}$, $G_{coh}^{(\sigma\bar{\sigma})}$ is the fluid-fluid cohesion strength, and it is used to tune the fluid-fluid interaction force. This parameter is used for tuning the surface tension in the pseudopotential model. It should be noted that only the inter-component interaction force is considered in this work ($\sigma \neq \bar{\sigma}$) while the intra-component interaction force is neglected $\left(G_{coh}^{(\sigma\sigma)} = 0\right)$. We set $G_{coh}^{(\sigma\bar{\sigma})} = 3.5$ in our simulations.

We point out that at boundary points where non-periodic boundary conditions are applied (e.g., Dirichlet boundary condition), there will be an imbalance in the force summation term of Eq. (13) as the neighbor points located at $(x + e_i)$ will be located outside the domain boundaries. In these cases, the missing terms are replaced with duplicates of the terms in the opposite direction ($\bar{\iota}$) (towards the interior of the domain instead of across the boundary of the domain) [37]. For example, at the left boundary of the domain, summation terms of pseudopotential in three directions would be missing, i.e., $i = 3,6,7$ as shown in Figure 1. The pseudopotential values in the neighbor lattice points adjacent to these points would be determined using as follows:

$$\psi^{(\bar{\sigma})}(x + e_3) = \psi^{(\bar{\sigma})}(x + e_1),$$
$$\psi^{(\bar{\sigma})}(x + e_6) = \psi^{(\bar{\sigma})}(x + e_8),$$
$$\psi^{(\bar{\sigma})}(x + e_7) = \psi^{(\bar{\sigma})}(x + e_5).$$

Same criterion was used in the right boundary of the domain for the missing terms. Such special treatment for interaction forces is essential with the implementation of boundary conditions mentioned in section 2.3.

The external body forces exerted on the $\sigma^{th}$ fluid component (e.g., gravity) could be added using the term $\boldsymbol{F}^{(\sigma),b}$. Several methods in the literature were introduced to incorporate these forces, one is referred to Ref. [11] for review of different techniques.

In the present pseudopotential model, the pressure is determined as follows:

$$P = c_s^2 \sum_\sigma \rho^{(\sigma)} + \frac{c_s^2 \Delta t}{2} G_{coh}^{(\sigma\bar{\sigma})} \sum_{\sigma,\bar{\sigma}} \rho^{(\sigma)} \rho^{(\bar{\sigma})}. \tag{14}$$

We express our results in lattice units (l.u.) in which length is represented in lattice length unit ($\Delta x = 1$), time is represented lattice time unit in ($\Delta t = 1$) and density is also the unit ($\rho = 1$). All other physical quantities, e.g., pressure, are also expressed in lattice units.

### 2.2. Improved virtual solid density scheme: wetting boundary condition

The fluid-solid interaction force $\boldsymbol{F}^{(\sigma),f-s}$ is defined as

$$\boldsymbol{F}^{(\sigma),f-s}(x) = -G_{ads}^{(\sigma)}\psi^{(\sigma)}(x) \sum_i w_i \, s^{(\bar{\sigma})}(x + e_i)e_i, \tag{15}$$

where, $G_{ads}^{(\sigma)}$ is the fluid-solid adhesion strength, and it is used to tune the fluid-solid interaction force (consequently contact angle $\theta$) in previous pseudopotential models. In this section, an alternative approach will be shown to tune the contact angle. $s^{(\bar{\sigma})}$ is the solid pseudopotential of the $\bar{\sigma}$ fluid component. In this work, the improved virtual solid density (IVSD) model [34] is used to define the solid pseudopotential as follows:

$$s^{(\sigma)}(x) = \phi(x)\, \tilde{\rho}^{(\sigma)}(x), \tag{16}$$

where, $\phi$ is a binary switch function which equals 0 for a fluid node and 1 for a solid node, $\tilde{\rho}^{(\sigma)}$ is the virtual solid density, and it is defined using the averaged density of the fluid as

$$\tilde{\rho}^{(\sigma)}(x) = \chi^{(\sigma)} \frac{\sum_i w_i \rho^{(\sigma)}(x + e_i)(1 - \phi(x + e_i))}{\sum_i w_i (1 - \phi(x + e_i))}. \tag{17}$$

$\chi^{(\sigma)}$ is a factor that controls the wettability of the $\bar{\sigma}$ fluid component. If the parameter $\chi^{(\sigma)}$ is unity, the fluid component is naturally wetting ($\theta = 90°$). To control the wettability of the fluid components, the parameter $\chi^{(\sigma)}$ is selected as follows: (only two components are considered in this work, i.e., $\sigma = 1, 2$)

$$\chi^{(1)} = 1 + \xi,$$
$$\chi^{(2)} = 1 - \xi,$$

where, fluid component-1 is forming the droplets and fluid component-2 is the surrounding fluid. Fluid component-1 becomes non-wetting if $\xi < 0$ and wetting if $\xi > 0$.

Practically, $G_{ads}^{(\sigma)}$ and $G_{ads}^{(\bar{\sigma})}$ are set equal to $G_{coh}^{(\sigma\bar{\sigma})}$ and the wettability is controlled by $\xi$. Hence, the same equation (Eq. 13) could be used for calculating the interaction forces using the neighboring pseudopotential values for fluid or solid.

It is beneficial to have a grouping of fluid wettability based on contact angle. In this work, the grouping shown in Table I [38] is used, and the contact angle is defined as in section (3.3).

*Table I. Definitions of wettability based on contact angle measurements.*

| Wettability state | Contact angle [º] |
|---|---|
| Complete water wet | 0 |
| Strongly water wet | 0 − 50 |
| Weakly water wet | 50 − 70 |
| Neutrally wet | 70 − 110 |
| Weakly oil-wet | 110 − 130 |
| Strongly oil-wet | 130 − 180 |
| Complete oil-wet | 180 |

### 2.3. Inlet and outlet boundary conditions

The definition and implementation of boundary conditions are critical to most fluid mechanics problems. It greatly affects the simulation and, if ill-defined, could lead to unphysical results [9]. Several boundary conditions were used in this work, for instance, periodic boundary conditions [9] and half-way bounce-back boundary conditions [9]. Simulations of realistic fluid problems would require the definition of additional types of boundary conditions at the inlet and outlet, e.g., to impose fixed pressure or velocity values. One technique to define a Dirichlet boundary condition is the well-known Zou-He method [32]. It is also known as the non-equilibrium bounce-back (NEBB) method [9]. This technique provides equations to determine the missing distribution functions at the boundary. In this work, we use it to define pressure boundary conditions. A complete set of equations used for inlet and outlet boundaries could be found in [32].

For simulation cases as in section 4, where pressure boundary conditions are needed at inlet and outlet sides, the known and missing distribution functions at boundaries for each component ($\sigma$) could be identified from Figure 1. In this work, we are concerned with two components only, hence $\sigma = 1,2$.

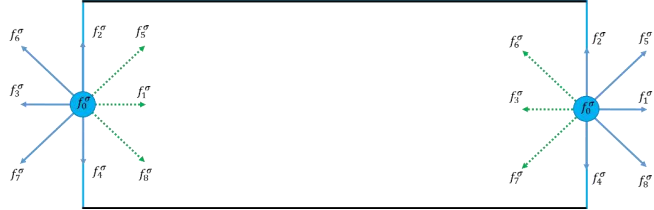

*Figure 1. Illustration of known and missing distribution functions at inlet and outlet boundaries. Solid blue lines represent known distribution functions, while dotted green lines represent unknown distribution functions.*

At the outlet side, there are (2) sets of (6) known distribution functions $\left( f_0^{(\sigma)}, f_1^{(\sigma)}, f_2^{(\sigma)}, f_4^{(\sigma)}, f_5^{(\sigma)}, f_8^{(\sigma)} \right)$ and their values are updated during the streaming step. Moreover, there are (2) sets of (3) missing distribution functions $\left( f_3^{(\sigma)}, f_6^{(\sigma)}, f_7^{(\sigma)} \right)$ as their values are not updated during the streaming step, and they would need information from outside the domain boundaries. Following the Zou-He technique to apply pressure boundary conditions, we assume zero normal velocity at the outlet. Moreover, we assume that fluid component-2 is the main fluid component present at the outlet side. Consequently, the following equations would define the first set of missing distribution functions at outlet [32],

$$u_x^{(2)} = -1 + \frac{f_0^{(2)} + f_2^{(2)} + f_4^{(2)} + 2 \left( f_1^{(2)} + f_5^{(2)} + f_8^{(2)} \right)}{\rho_{out}^{(2)}}, \tag{18}$$

$$f_3^{(2)} = f_1^{(2)} - \frac{2}{3} \rho_{out}^{(2)} u_x^{(2)}, \tag{19}$$

$$f_6^{(2)} = f_8^{(2)} - \frac{1}{2} \left( f_2^{(2)} - f_4^{(2)} \right) - \frac{1}{6} \left( \rho_{out}^{(2)} u_x^{(2)} \right), \tag{20}$$

$$f_7^{(2)} = f_5^{(2)} + \frac{1}{2} \left( f_2^{(2)} - f_4^{(2)} \right) - \frac{1}{6} \left( \rho_{out}^{(2)} u_x^{(2)} \right), \tag{21}$$

where, $\rho_{out}^{(2)}$ is the specified density of fluid component-2 at the outlet boundary, and it is used to impose an outlet pressure boundary condition.

Similarly, the following equations would define the first set of missing distribution functions at inlet [32],

$$u_x^{(1)} = 1 - \frac{f_0^{(1)} + f_2^{(1)} + f_4^{(1)} + 2 \left( f_3^{(1)} + f_6^{(1)} + f_7^{(1)} \right)}{\rho_{in}^{(1)}}, \tag{22}$$

$$f_1^{(1)} = f_3^{(1)} + \frac{2}{3}\,\rho_{in}^{(1)} u_x^{(1)}, \tag{23}$$

$$f_5^{(1)} = f_7^{(1)} - \frac{1}{2}\left(f_2^{(1)} - f_4^{(1)}\right) + \frac{1}{6}\left(\rho_{in}^{(1)} u_x^{(1)}\right), \tag{24}$$

$$f_8^{(1)} = f_6^{(1)} + \frac{1}{2}\left(f_2^{(1)} - f_4^{(1)}\right) + \frac{1}{6}\left(\rho_{in}^{(1)} u_x^{(1)}\right), \tag{25}$$

where, $\rho_{in}^{(1)}$ is the specified density of fluid component-1 at the inlet boundary, and it is used to impose an inlet pressure boundary condition.

Until this stage, the second set of missing distribution functions at the outlet $\left(f_3^{(1)}, f_6^{(1)}, f_7^{(1)}\right)$ and inlet $\left(f_1^{(2)}, f_5^{(2)}, f_8^{(2)}\right)$ would still be missing. We identify two techniques from the literature to define the sets of missing distribution functions. The first one would be to define the missing values by assigning a fixed density for the other fluid component as well [20]. Hence Eqs. (18-21) and (22-25) would be applied to fluid component-2 and fluid component-1, respectively, after altering the subscripts representing the fluid component. We call this treatment of boundary conditions, **Set-1**. The second technique would be to define the missing values using the method described in Ref. [33] where the unknown distribution functions maintain themselves in the update process as,

$$f_{3,6,7}^{(1)}\big|_{out,t_1} \xrightarrow{collision} f_{3,6,7}^{(1)}\big|_{out,t_2} \xrightarrow{streaming} f_{3,6,7}^{(1)}\big|_{out,t_1}, \tag{26}$$

$$f_{1,5,8}^{(2)}\big|_{in,t_1} \xrightarrow{collision} f_{1,5,8}^{(2)}\big|_{in,t_2} \xrightarrow{streaming} f_{1,5,8}^{(2)}\big|_{in,t_1}, \tag{27}$$

where, $t_1$ is the current time step and $t_2$ is the new time step. We call this treatment of boundary conditions, **Set-2**. It will be shown in section 4.1 that applying pressure boundary conditions using either **Set-1** or **Set-2** resulted in a considerable change in the volume of the trapped fluid component within the simulation domain. This is an unphysical feature that would affect the simulation of displacement processes in porous media where entrapment of fluid components is expected to be encountered.

In the following paragraphs, we introduce a newly developed set of treatments for inlet and outlet pressure boundary conditions that will be shown to suppress this non-physical behavior. We call the newly developed treatments **Set-3**, and they are implemented as follows:

1- Apply Zou-He technique for only one fluid component at each side as shown in Eqs. (18-21) and (22-25).

2- Apply Eq. (26) as an open boundary condition for fluid component-1 at the outlet.

3- The set of missing distribution functions of fluid component-2 at the inlet will be updated during the collision step – by applying Eq. (1) – and will not be updated during the streaming step. Hence, they will keep their post-collision values as follows:

$$f_{1,5,8}^{(2)}\big|_{in,t_1} \xrightarrow{collision} f_{1,5,8}^{(2)}\big|_{in,t_2} \xrightarrow{streaming} f_{1,5,8}^{(2)}\big|_{in,t_2}. \tag{28}$$

These three steps represent **Set-3** of treatments for boundary conditions. This set combined with the modified EF model Eq. (1), special treatment for interaction forces at boundaries, and the IVSD model Eqs. (16,17) represents the improved MC pseudopotential model.

It should be noted that in **Set-3,** the pressure value at boundaries will not be altered only by the input density at boundaries since the other density component is freely updated and will be calculated using Eq. (3). Hence, according to the model EOS (Eq. 14), the pressure value will not be fixed at both boundaries, and it shall be evaluated using the EOS to determine the exact value. In this work, we report the average pressure values, which are determined over the boundaries length, and ignore the pressure values near the solid walls. Moreover, step-2 is applied to the fluid component with higher kinematic viscosity, which is assumed to be fluid component-1 in this work, while step-3 is applied to the fluid component with lower kinematic viscosity, which is assumed to be fluid component-2 in this work. Velocity artifacts near the boundaries were observed in case of reversing the component to which the relevant step was applied.

## 3. NUMERICAL TESTS

In this section, we performed benchmark tests to validate and characterize the developed model. Some of these benchmarks were also used to calibrate the model parameters which were needed for other simulations, such as surface tension and contact angle.

### 3.1. Miscibility test

This numerical test is one of the standard tests for the MC pseudopotential LBM model that could be used to determine the effect of $G_{coh}^{(\sigma\bar{\sigma})}$ on the densities of the fluid components. It also shows the dependence of the density and surface tension values on the viscosity ratio of the two components. In this section, we carry out the test at three values of viscosity ratio $\left( M = 1, M = 25, M = \frac{1}{25} \right)$ to show the effect of viscosity on the results. The relaxation times to achieve these viscosity ratios were set as follows:

$$\tau^{(\sigma)} = \begin{cases} \tau^{(1)} = \tau^{(2)} = 1, & M = 1 \\ \tau^{(1)} = 3.5, \tau^{(2)} = 0.62, & M = 25 \\ \tau^{(1)} = 0.62, \tau^{(2)} = 3.5, & M = \frac{1}{25}. \end{cases}$$

Series of 2D simulations were carried out where a droplet of fluid component-1 was placed in a fully periodic domain of fluid component-2. The domain size was 192 x 192, and the initial droplet size was 40. The density field was initialized using the following hyperbolic formula

$$\rho^{(\sigma)}(\mathbf{x}) = \frac{\rho_{in}^{(\sigma)} + \rho_{out}^{(\sigma)}}{2} - \frac{\rho_{in}^{(\sigma)} - \rho_{out}^{(\sigma)}}{2} \times \tanh\left[ \frac{2\left( \sqrt{(x - x_c)^2 + (y - y_c)^2} - r_0 \right)}{W} \right], \quad (29)$$

where, $\rho_{in}^{(\sigma)} = 1$, $\rho_{out}^{(\sigma)} = 0.02$, $r_0 = 40$ and $W = 5$. $x_c$ and $y_c$ are locations of the domain's center. This initialization method reduces the numerical instability in the first few iterations of the simulation as it represents an initial density field with a diffusive interface between the fluid components. It should also be noted that starting the simulation with an initial density $\rho_{out}^{(\sigma)}$ far from the expected equilibrium value results in a longer convergence time and a dramatic change in the droplet size as the simulation converges at a different droplet size. Hence, the initial dissolved density was set to 0.02, a value obtained from several trials and expected to be close to the equilibrium dissolved density value.

After the simulation reached a steady state, the density values of fluid component-1 and fluid component-2 at the center of the droplet were measured and shown in Figure 2. Also, the pressure value at the center of the droplet and the domain corner away from the droplet were measured and used to calculate the surface tension using Laplace law (section 3.2).

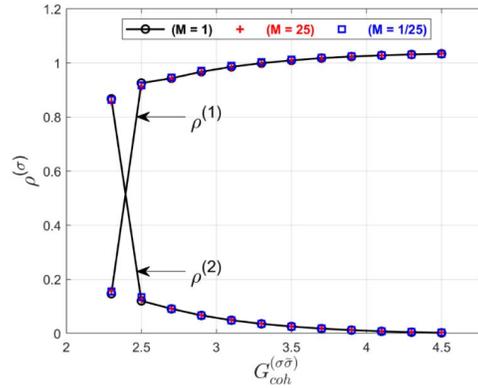

*Figure 2. The density of the two fluid components in the miscibility test. The left axis shows the component density measured at the center of the domain, while the bottom axis shows the selected value of $G_{coh}^{(\sigma\bar{\sigma})}$. Different symbols represent different viscosity ratios.*

It could be observed from Figure 2 that as the value of $G_{coh}^{(\sigma\bar{\sigma})}$ was increased, the density difference between the suspended fluid (fluid component-1) and the suspending fluid (fluid component-2) was increased, which indicates higher segregation between the two fluid components. Also, we noted from Figure 3 that the surface tension increased with $G_{coh}^{(\sigma\bar{\sigma})}$. Moreover, there was a critical value of $G_{coh}^{(\sigma\bar{\sigma})}$ below which the cohesion interaction strength was not strong enough to maintain a defined interface between the two fluid components, and they became completely miscible. This critical value could be identified from Figure 2 as $G_{coh}^{(\sigma\bar{\sigma})} < 2.5$ below which the density of fluid component-1 at the center is greatly reduced and the density of the fluid component-2 is significantly increased, indicating miscibility of the first fluid component in the second fluid component.

Finally, we observed that changing the viscosity ratio did not result in major changes in the density and surface tension values.

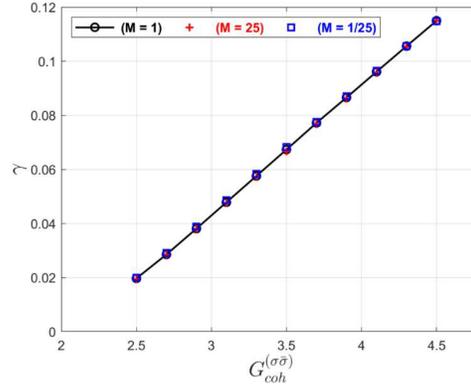

*Figure 3. Surface tension in the Laplace test. The left axis shows the surface tension, while the bottom axis shows the selected value of $G_{coh}^{(\sigma\bar{\sigma})}$. Different symbols represent different viscosity ratios.*

### 3.2. Laplace test

This test is carried out to verify that the model obeys the Laplace law $\left(\Delta P = \frac{\gamma}{R}\right)$, where $\Delta P$ is the difference between the pressure inside the droplet and the pressure outside the droplet, $\gamma$ is the surface tension, and $R$ is the droplet radius. Laplace law states that there is a linear relationship between the pressure difference and the inverse of the droplet radius, with the constant of this linear relationship being the surface tension.

We carried out 2D simulations with a droplet of fluid component-1 placed in a fully periodic domain of fluid component-2. The domain size was 192 x 192, and the initial droplet size was 40. The density field was initialized using Eq. 29. The initial density values were selected as follows (from miscibility test at $G_{coh}^{(\sigma\bar{\sigma})} = 3.5$): $\rho_{in}^{(1)} = \rho_{out}^{(2)} = 1$, $\rho_{out}^{(1)} = \rho_{in}^{(2)} = 0.027$, such values are selected to reduce the change in the droplet radius from the initial radius as the simulation converges to a steady state. Three different viscosity ratios were simulated $\left(M = 1, M = 25, M = \frac{1}{25}\right)$ by setting the relaxation times as shown in the miscibility test (section 3.1).

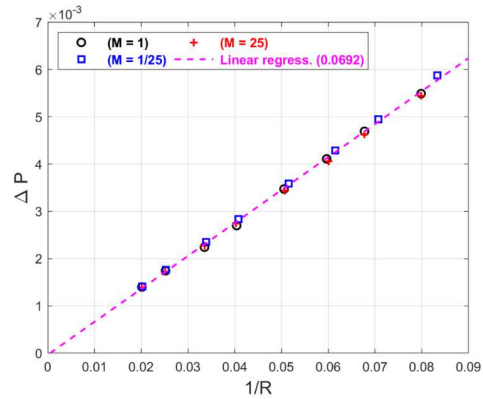

*Figure 4. Laplace test results. The left axis shows the pressure difference, while the bottom axis shows the inverse of the droplet radius. Different symbols represent different viscosity ratio (black circle: M = 1, red cross: M = 25, blue square: M = 1/25). The dotted line in magenta represents the linear regression of data which gives a surface tension (slope) value $\gamma = 0.0692$ ($G_{coh}^{(\sigma\bar{\sigma})} = 3.5$).*

After the simulations reached a steady state, the pressure inside and outside the droplet was measured and plotted against the inverse of the measured droplet radius as shown in Figure 4. The relationship between $\Delta P$ and $R$ was indeed linear and thus satisfied the Laplace law. The surface tension was determined by the slope of the line. It was shown in Ref. [39] that the relationship between pressure difference ($\Delta P$) and curvature $\left(\frac{1}{R}\right)$ depends on the selected viscosity ratio in the original pseudopotential model, which is one of its limitations. However, the present model does not suffer from this problem. As shown in Figure 4 that the surface tension does not depend on the selected viscosity ratio. Results in Figure 3 and Figure 4 show that the present model exhibits independent tuning surface tension from the viscosity ratio.

### 3.3. Contact angle measurement

This benchmark is conducted to calibrate the wetting boundary condition using the (IVSD) scheme introduced earlier (section 2.2) and determine the relationship between the parameter $\xi$ and contact angle $\theta$ for the currently implemented model. In the pseudopotential model, the contact angle cannot be set directly into the model like, for instance, in the free energy model [9]. Hence, this characterization is essential to calibrate the model at a specific surface tension $\left(G_{coh}^{(\sigma\bar{\sigma})}\right)$ and properly set the desired contact angle in other simulations.

We carried out 2D simulations where a semi-circular droplet of fluid component-1 with an initial droplet radius of 30 was initially placed on the surface of a solid wall. The domain size was 192 x 96 and periodic conditions in the x-direction, while solid walls were located at the top and the bottom of the domain. The initial density values and $G_{coh}^{(\sigma\bar{\sigma})}$ were set like in section 3.2 using Eq. 29. Three different viscosity ratios were simulated $\left(M = 1, M = 25, M = \frac{1}{25}\right)$ by setting the relaxation times as shown in section 3.1.

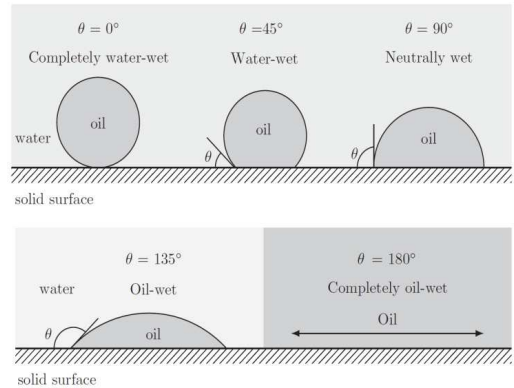

*Figure 5. An illustration of different contact angles measured through water in an oil/water system [2]. Here a droplet of oil is placed on a solid mineral surface, surrounded by water. The contact angle is measured through the water.*

We specify the criteria for measuring the contact angle ($\theta$) by the convention that the contact angle is measured through the denser fluid phase [2]. Since this work focused on an oil-water system, water is considered the denser fluid, and contact angle will be measured through the water, as shown in Figure 5 [2]. The initial density field was initialized with the assumption that fluid component-1 (forming the droplet) is oil while fluid component-2 (surrounding fluid) is water. The contact angle of the droplet was measured in the steady state using the technique and geometrical relationships shown in Ref. [40], where the base width was measured five lattice units above the solid surface to avoid the influence of the surface on the measured interface. Additionally, a linear interpolation was used to determine the approximate location of the interface. Also, the half-way location of solid walls was considered [9].

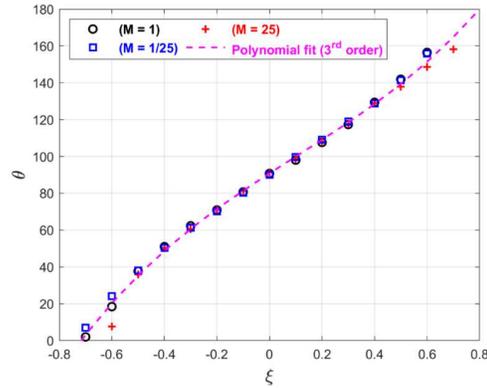

*Figure 6. Contact angle test. The left axis shows the contact angle θ, while the bottom axis shows the factor ξ. Different symbols represent different viscosity ratios. The dotted line in magenta shows a polynomial (3rd order) data fit.*

Figure 6 depicts the relationship between $\xi$ and contact angle $\theta$. We noted that the model produced the expected behavior in setting the wettability using the parameter $\xi$, where oil was non-wetting ($\theta < 90°$) for $\xi < 1$, and vice versa. Also, a wide range of contact angles was successfully simulated, covering most practical applications. Additionally, the contact angle for the selected $\xi$ did not change considerably with the viscosity ratio, except at the extremes of the curve, so the chosen contact angle is independent of the viscosity ratio with attention to the extreme cases (e.g., superhydrophobic surfaces). Moreover, an empirical equation could be deduced from the curves to represent the $\xi$-$\theta$ relationship: a third-order polynomial provided the best fit for our simulation results ($\theta = 47.5487\,\xi^3 - 13.9067\,\xi^2 + 92.0987\,\xi + 90.8074$). Snapshots of the simulation results at different contact angles are shown in Figure 7.

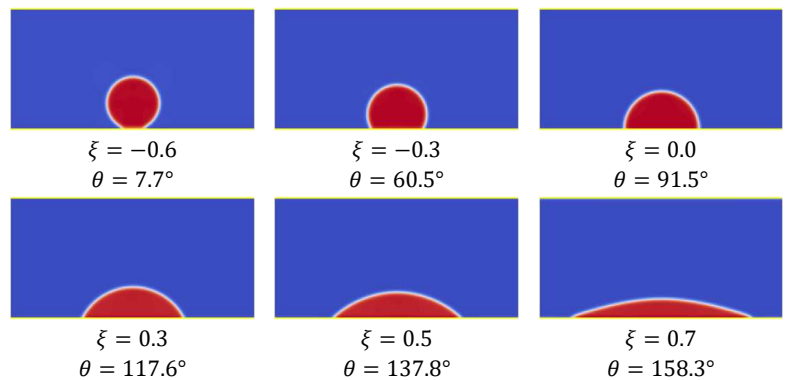

$\xi = -0.6$          $\xi = -0.3$          $\xi = 0.0$
$\theta = 7.7°$          $\theta = 60.5°$          $\theta = 91.5°$

$\xi = 0.3$          $\xi = 0.5$          $\xi = 0.7$
$\theta = 117.6°$          $\theta = 137.8°$          $\theta = 158.3°$

*Figure 7. Snapshots of contact angle test at different values of ξ with corresponding values of contact angle θ. The contact angle is measured through the water phase (surrounding fluid).*

## 4. Results

### 4.1. Fluid flow over an obstacle with entrapped fluid volume

In this section, we evaluate the performance of the presented model in simulating immiscible fluid displacement. We check several features that make the model superior to previous pseudopotential models. For instance, capturing physical phenomena and retaining a fixed volume of entrapped fluid components are vital characteristics. This is essential as previous pseudopotential models were shown to experience changes in trapped fluid volumes for similar simulations [28].

We simulated a simple case in which a semi-circular droplet of fluid component-1 was placed on the vertical side of a u-shaped solid wall surrounded by fluid component-2. Moreover, fluid component-1 was pushed inside the fluid channel using pressure boundary conditions. Such conditions were set to test behaviors of trapped volumes of the different fluid components under pressure boundary conditions and to evaluate if there were changes in the set contact angle under such conditions.

The setup of our 2D simulations is shown in Figure 8, where an 800 x 200 flow channel was partially filled with fluid component-2. The channel contained a u-shaped wall that served as an obstacle to the flow and would force the flow of fluid component-1 to the narrower upper and lower fluid passages, and it would also result in entrapment of fluid component-2 by these upper and lower flow streams. The initial droplet radius was 30. Fluid component-1 (invading fluid) was set to be more viscous and non-wetting. Hence the simulation parameters were set as follows: $\tau^{(1)} = 3.5 \ (\nu^{(1)} = 1), \tau^{(2)} = 0.62 \ (\nu^{(2)} = 0.04), M = 25, \xi = -0.4 (\theta \approx 50°)$ and the fluid-fluid cohesion strength was set as $G_{coh}^{(\sigma\bar{\sigma})} = 3.5$.

The set of boundary conditions described in section 2.3 was used, where the inlet density (pressure) of fluid component-1 was set to 1.01, and the outlet density (pressure) of fluid component-2 was set to 1. The results of the three boundary condition treatments were compared as shown in Figure 8. It could be observed that in all three sets of boundary conditions, there was no reduction in the volume of the non-wetting fluid droplet on the vertical side of the u-shaped wall, where the different sets of boundary conditions were combined with the IVSD wetting boundary conditions. Moreover, we observed that set-1 and set-2 resulted in a considerable change in the volume of the trapped fluid component due to the passage of the invading fluid (red) through the original fluid (blue), which is non-physical. However, set-3 greatly suppressed that unphysical behavior, and the entrapped fluid component survived until the end of the simulation. Preservation of such trapped fluid volumes is essential for the simulations of immiscible displacement of fluids in porous media.

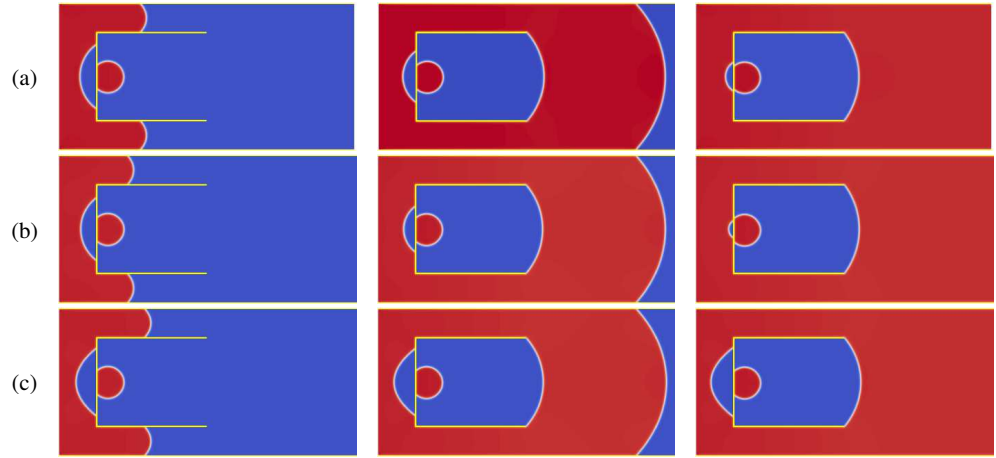

*Figure 8. Comparison between simulation results for the displacement of immiscible fluids in a channel and over a u-shaped wall using different sets of boundary conditions: (a) Set-1, (b) Set-2, and (c) Set-3. Snapshots are shown at different time steps and for the first half of the channel length (400 x 200). The colors represent fluid component-1 density field where red represents higher density (fluid component-1) and blue represents lower density (fluid component-2), with the interface region represented with the gradient of colors. The top, bottom, and u-shaped solid walls are shown in yellow. Fluid component (1) invades the domain from the left side. Snapshots are ordered from left to right in each row for a different set of boundary condition treatments.*

The evolution of the fluid invasion using set-3 of boundary condition treatments is shown in Figure 9 via snapshots at different time steps. Since the invading fluid is non-wetting, at t =10000, it first started to invade the regions away from the upper and solid walls. However, it was resisted by fluid component-2 in the central part where the u-shaped wall was located. Also, by comparing snapshots at t = 0 and t = 10000, it could be seen that the droplet of fluid component-1 initially set as neutrally wet changed to a non-wetting droplet. The measured contact angle was θ ≈ 50° as initially set using ξ = −0.4. Another interesting phenomenon could be observed in the upper and lower passages, where one side of the invading fluid is preceding the other. Since the invading fluid is non-wetting, it started invading regions away from the solid walls firstly, and by the time it approached the flow passages, the side of the fluid next to the solid channel wall was already succeeding other fluid regions. This matches the expected physical behavior of the non-wetting fluid in such cases. The invading fluid was forced to flow through the upper and lower flow passages, where it was sandwiched by two solid walls as shown in the snapshot at t = 30000. The entrapped volume of fluid component-2 could be observed throughout the remaining snapshots until they finished, and it kept its shape without considerable changes.

The coalescing process of the upper and lower portions of the invading fluid with the form of a single interface of the invading fluid could be observed in the snapshots at t = 205000 and t

= 210000. The non-wetting fluid kept invading the channel until it successfully reached the channel's end, as shown in the last snapshot at t = 700000. It could also be observed that a film of fluid-component-2 remained in the channel since a fixed density (pressure) boundary condition is specified at the outlet. The entrapped volumes of both fluid components kept their shape and size till the end of the simulation. It should be mentioned that there was a slight reduction in the entrapped volume of fluid component-2 inside the U-shaped wall that occurred as the invading fluid reached the end of the fluid channel. It is believed that such behavior occurred as the pressure inside the channel was increased due to the blockage at the end of the channel. Also, the LBM model used in this work is considered weakly compressible, and that would allow small changes in the fluid volumes.

The simulation results in this section showed that the presented LBM model suppressed the change in volumes of trapped fluid components, produced stable contact angles with pressure boundary conditions, and captured the underlying physics.

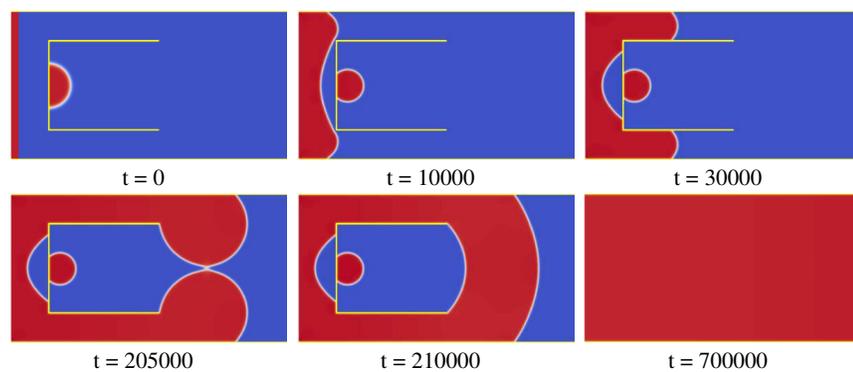

| t = 0 | t = 10000 | t = 30000 |
| t = 205000 | t = 210000 | t = 700000 |

*Figure 9. Simulation results for the displacement of immiscible fluids in a channel and over a u-shaped wall using the set-3 for the boundary condition. Snapshots are shown at different time steps and for half the channel, starting at initial conditions ($t = 0$) (top left snapshot) until the invading fluid reached the end of the channel (bottom right snapshot). All snapshots are taken for the first half of the channel where the u-shaped is located, except for the last snapshot at t = 700000, which is taken for the second half of the channel to show the arrival of the invading fluid at the outlet boundary. The colors represent fluid component-1 density contour field where red represents higher density (fluid component-1) and blue represents lower density (fluid component-2) with the interface region represented with a gradient of colors.*

### 4.2. Flow in heterogeneous porous media

#### 4.2.1. Primary Drainage

In this section, we simulate fluid displacements processes when one fluid replaces another in a porous medium. The initial state of the porous medium is assumed to be a clean and saturated porous medium where the solid walls are water wet. This matches the initial state of laboratory experiments carried out on porous samples for oil reservoirs where the extracted samples (core plugs) are cleaned and fully saturated with formation water. Starting with the pore space entirely saturated with water (wetting fluid) and pushing oil (non-wetting fluid) into the pore space is known as the primary drainage process [2]. The word primary (meaning first) indicates that this is the first time the non-wetting phase enters the pore space. We start with a wetting phase saturation of $S_w = 1$. Drainage, in general, refers to the displacement of a wetting phase by a nonwetting phase or decreasing the wetting phase saturation. A typical example of a primary drainage process is oil migration from source rock to an oil reservoir where oil invades the pores initially saturated with water.

In laboratory experiments, the non-wetting phase does not penetrate the medium until the capillary pressure exceeds a threshold pressure $P_{c,t}$, which depends on the size and shape of the pores and wettability of the sample [1]. As the capillary pressure increases beyond this value, the saturation of water continues to decrease, with water saturation approaching an irreducible level $S_{w,i}$ at very high capillary pressures. Irreducible water saturation $S_{w,i}$ is the lowest water saturation that could be achieved by a displacement process [1].

The geometry we used in our simulations was adopted from previous work [24]. We briefly describe the geometry here. The generated geometry is bounded by upper and lower solid walls and filled with solid blocks creating the porous medium. The size of each solid block is 10 x 10, while the size of the whole domain is 400 x 200 lattices. The domain porosity is 81.17%. Internal solid blocks were removed from the first and last ten layers of the inlet and outlet sides to represent reservoirs for both fluids. Five layers of oil are assumed in the assigned inlet fluid reservoir. The initial density values are as assumed in section 3.2, i.e., $\rho_i^{(1)} = 1.0$ , $\rho_i^{(2)} = 0.027$ where the layers of the non-wetting fluid are located. Other domain regions are assumed to have similar but reversed initial density values. The virtual solid density values were computed at the initial state using the technique mentioned in section 2.2, and the required wettability condition dedicates their values.

In these simulations, the non-wetting fluid (oil) is set to be more viscous than the wetting fluid (water). A realistic viscosity ratio was chosen as $M = \frac{\rho_{nw}\nu_{nw}}{\rho_w\nu_w} = 25$, where the density of both fluids was set as unity. In this length scale (pore-scale $- \mu m$), the density difference is not expected to have a major impact on the displacement process, and in realistic experiments, the sample is mostly placed horizontally. Hence the gravity effects could be neglected. The relaxation times are set as in section 3.1. The contact angle was varied to represent different cases and inspect the effect of the contact angle ranging from strongly water-wet to oil-wet mediums. In this section, the contact angle was set as $\theta \cong 50°\,(\xi = -0.4)$ to represent a weakly water-wet medium.

The capillary pressure $P_c$ is considered the difference between the inlet and outlet pressures $P_c = P_{in} - P_{out}$. Inlet and outlet pressures are specified using Dirichlet boundary conditions via the Zou-He technique as explained in section 2.2, and set-3 was used in these simulations.

Inlet and outlet density values of the relevant fluid component were altered to employ the pressure boundary conditions. Despite the Zou-He method being a common technique to apply pressure boundary conditions in the LBM, it should be expected to have a negative impact on multicomponent simulations due to the inherent compressibility effects of the model. For the primary drainage simulation, the outlet density was kept fixed at $\rho^{(2)} = 1$ while the inlet density increased gradually and stabilized after some timesteps. Water saturation ($S_w$) was determined as the number of lattice points occupied by water divided by the total number of fluid lattice points.

The primary drainage simulation results are plotted in Figure 10. It could be observed that the current LBM model successfully captures the primary drainage curve in a qualitative manner. The oil started invading the medium after exceeding a critical threshold pressure ($P_{c,t} \cong 0.00511$) at a critical oil saturation $S_{o,c} \cong 0.053$ and kept invading the medium gradually with the increase of capillary pressure. An interesting phenomenon occurred at $S_w \cong 0.7$ where the capillary pressure was considerably increased, but the water saturation was not decreased. This is known as the capillary pressure bump phenomenon, and it is known to occur due to the heterogeneity of the porous medium [1]. The LBM was first shown to capture this phenomenon in Ref. [24] using the original pseudopotential model, and it is reassured in this study to be captured using the present model. Another smaller bump in the capillary pressure curve was also captured using the present model – for the same geometry – at $S_w \cong 0.35$. The oil kept invading the medium with the gradual increase of the capillary pressure until $P_c \cong 0.01136$ and $S_w \cong 0.2$ where the oil arrived at the wetting fluid reservoir at outlet. Moreover, any additional increase in the capillary pressure did not result in any further decrease/increase in the water/oil saturation. This indicates that the irreducible water saturation is achieved at $S_{w,i} \cong 0.2$, which is close to realistic values of $S_{w,i}$ [1].

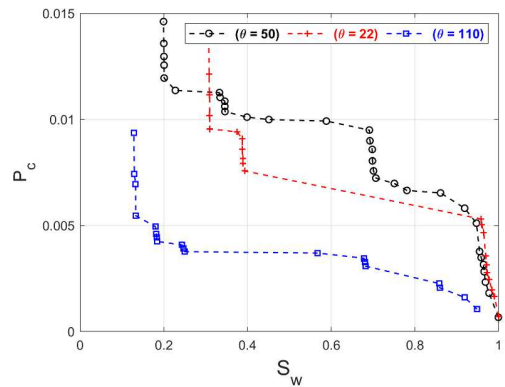

*Figure 10. Capillary pressure ($P_c$) – water saturation ($S_w$) relationships for the primary drainage simulation process at different contact angles and viscosity ratio $M = 25$. Black circles represent the results at $\theta \cong 50°$ (weakly water wet), red crosses represent the results*

*at $\theta \cong 22°$ (strongly water-wet), and blue squares represent the results at $\theta \cong 110°$ (weakly oil-wet).*

It should be highlighted that such realistic values of $S_{w,i}$ is one of the main advantages of the present model, as the proposed treatments and boundary conditions greatly suppressed the shrinkage of the entrapped fluid volume, which is a known drawback of the pseudopotential model [24] [28]. For instance, Ref. [24] used the same geometry as in the present study, but the entrapped volumes of water shrank, resulting in non-realistically low irreducible water saturations. However, using the present improved model, the trapped water volumes survived until the end of the primary drainage process with the increase of the inlet pressure. Notice that relatively small volumes of water did not survive and collapsed. This is inevitable for the pseudopotential model as it is a diffusive interface model. Hence, approaching an interface width that is close to the fluid volume will result in such collapses. Moreover, some trapped water volumes experienced a moderate change in volume with increased capillary pressure, which is due to the compressibility effects of this model. Snapshots of the density distribution (to represent fluids distribution) in the porous medium are shown in Figure 11 (a) for the weakly water-wet system ($\theta \cong 50°$).

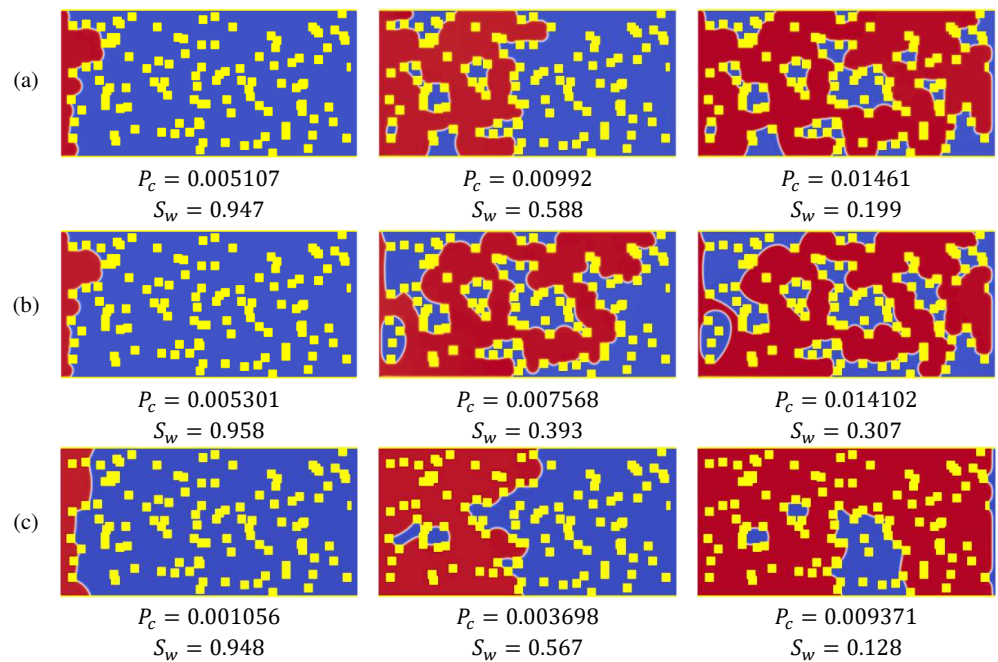

(a)

$P_c = 0.005107$    $P_c = 0.00992$    $P_c = 0.01461$
$S_w = 0.947$       $S_w = 0.588$      $S_w = 0.199$

(b)

$P_c = 0.005301$    $P_c = 0.007568$   $P_c = 0.014102$
$S_w = 0.958$       $S_w = 0.393$      $S_w = 0.307$

(c)

$P_c = 0.001056$    $P_c = 0.003698$   $P_c = 0.009371$
$S_w = 0.948$       $S_w = 0.567$      $S_w = 0.128$

*Figure 11: Snapshots for the primary drainage process at different wettability conditions: (a) weakly water-wet medium $\theta \cong 50°$, (b) strongly water-wet medium $\theta \cong 22°$ and (c) weakly oil-wet medium $\theta \cong 110°$. The Snapshots were taken after oil stopped invading the medium*

*at the specified capillary pressure. Oil invaded the medium from the left side. Snapshots are ordered from left to right in each row.*

In the following sections, the same simulation parameters were set as in the previous section. However, the contact angle was changed to represent different medium wettability states.

### 4.2.1.1. Strongly water wet medium

In this case, the contact angle was set as $\theta \cong 22°$ ($\xi = -0.55$) to represent a strongly water-wet system. This would represent a core plug after being cleaned in the laboratory and fully saturated with water where the sample becomes strongly water-wet.

The primary drainage results of this case are shown in Figure 10. Interesting changes were observed compared to the weakly water-wet case in the previous section. The critical threshold pressure slightly increased in this case ($P_{c,t} \cong 0.005301$). Despite that increase is relatively small, it matches the expected physical behavior as the oil is less wetting in this case. Hence, it would need a larger capillary pressure threshold to start invading the medium.

From the snapshots of the time evolution of the displacement process in Figure 11 (b), we observed that the displacement pattern changed. Consequently, the capillary pressure bump did not occur in the same region as in the previous case. However, it occurred in a different region at $S_w \cong 0.388$. Wettability is known to alter the shape of the capillary pressure curves [1], and such a change in the simulation results shows that the current LBM model can capture these effects. Since the oil phase is less wetting in this case, it had less tendency to stick to solid walls and more tendency to occupy pores away from the solid walls, which resulted in the high increase in the oil saturation once the capillary pressure exceeded the critical threshold pressure. In this case, oil was able to invade more pore spaces at lower capillary pressures due to its preference to fill pores away from solid walls. The nature of the used geometry aided in such behavior as it had relatively large pores. Moreover, it was observed that the irreducible water saturation $S_{w,i}$ was increased in this case to $S_{w,i} \cong 0.3$ which matches the expected physical behavior. As the medium had a stronger tendency for water wetting, more water was trapped in pore spaces that had larger interfaces with solid walls.

### 4.2.1.2. Weakly oil-wet medium

In this case, the contact angle was set as $\theta \cong 110°$ ($\xi = 0.2$) to represent a weakly oil-wet system. This would represent a typical oil reservoir where the porous medium is oil-wet rather than water-wet [1]. This case was called water drainage to avoid contradiction with the definition of drainage.

The results of this case are shown in Figure 10. We observed that the critical threshold pressure decreased, and oil immediately started invading the medium. That matches the expected physical behavior since oil was the wetting phase in this case and had the tendency to stick to solid walls. Again, by looking at the displacement snapshots in Figure 11 (c), we observed that the displacement pattern changed, and the wetting fluid is invading more pore spaces with less non-wetting trapped. Also, the capillary pressure bump phenomenon was suppressed as the wetting fluid is less resistant to flow in the medium. Moreover, the irreducible water saturation $S_{w,i}$ decreased in this case to $S_{w,i} \cong 0.128$ which matches the expected physical behavior. As the medium had a stronger tendency for oil wetting, less water was trapped in pore spaces that had larger interfaces with solid walls.

It could be concluded from this section that the presented model captured different physical behaviors and phenomena that occur during the primary drainage displacement process in a qualitative manner.

### 4.2.2. Imbibition

After completing simulations of capillary pressure for primary drainage, the direction of saturation change can be reversed, and another capillary pressure relationship can be measured. It is usually called an imbibition relationship. Imbibition, in general, refers to the displacement of a non-wetting phase by a wetting phase or the increase of wetting phase saturation. A common example of imbibition is waterflooding process, where water is injected into an oil reservoir to displace oil. The primary drainage and imbibition relationships differ considerably. This difference is called capillary pressure hysteresis [1].

A series of 2D simulations were carried out to simulate the imbibition process in the same medium used in the previous section. The endpoints of the primary drainage curves were used as the initial saturation of the imbibition simulations. The capillary pressure was changed by reducing the inlet density of fluid component-1 and fixing the outlet density of fluid component-2 until $P_c = 0$ to capture the spontaneous imbibition phenomenon, then an increase in the density of fluid component-2 resulted in negative capillary pressures to simulate forced imbibition. Also, we use the same contact angle as in the primary drainage case $\theta \cong 50° \ (\xi = -0.4)$.

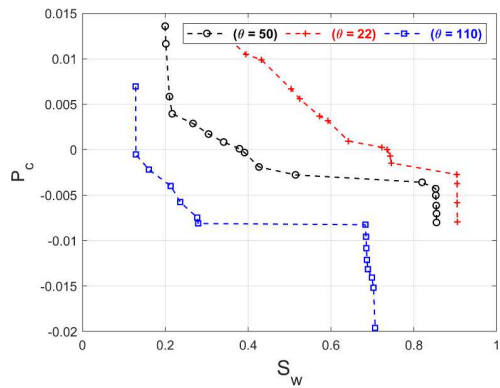

*Figure 12. Capillary pressure ($P_c$) – water saturation ($S_w$) relationships for the imbibition simulation process at different contact angles and viscosity ratio M = 25. Black circles*

*represent the results at $\theta \cong 50°$ (weakly water wet), red crosses represent the results at $\theta \cong 22°$ (strongly water-wet), and blue squares represent the results at $\theta \cong 110°$ (weakly oil-wet).*

The imbibition simulation results are plotted in Figure 12, where it could be observed that for the initial reductions of capillary pressure, the water saturation did not considerably increase. However, as the capillary pressure approached zero, there was an increase in the water saturation, which shows that the model captured the spontaneous imbibition phenomenon. Also, the water saturation at zero capillary pressure is not very large, which matches the expected physical behavior when the capillary pressure reaches zero at a lower saturation for a less strongly wetting phase [1].

Snapshots for the imbibition process are shown in Figure 13 (a). The oil started retracting from the medium, gradually increasing the water saturation until the medium was mostly filled with water. The last snapshot shows trapped volumes of oil that are known as residual oil. The residual oil saturation, in this case, was $S_{o,r} = 0.145$, which is relatively small for a weakly water-wet medium [1]. However, this could be attributed to the nature of the simulated geometry, which has rather large porosity and pore spaces.

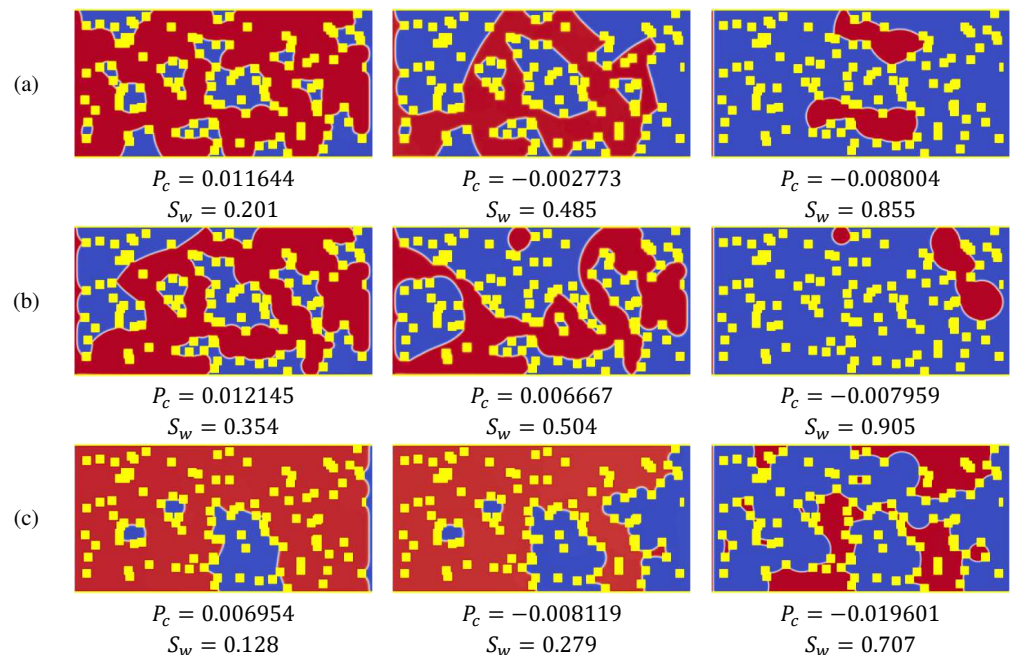

*Figure 13 Snapshots for the imbibition process at different wettability conditions: (a) weakly water-wet medium $\theta \cong 50°$, (b) strongly water-wet medium $\theta \cong 22°$ and (c) weakly oil-wet medium $\theta \cong 110°$. The Snapshots were taken after the water stopped invading the medium at the specified capillary pressure. Water invaded the medium from the right side. Snapshots are ordered from left to right in each row."*

In the following sections, the same simulation parameters were set as in the previous section of primary drainage simulations, where the contact angles were altered to represent different wettability states. As mentioned earlier, the endpoints of the different primary drainage curves were used as initial saturation for the imbibition simulations.

### 4.2.2.1. Strongly water wet medium

In this case, the medium was set as strongly water wet with a contact angle as $\theta \cong 22°$ ($\xi = -0.55$) during the primary drainage process, where oil was observed to need a higher capillary pressure to invade the medium and irreducible water saturation was also relatively high.

The imbibition process results are shown in Figure 12, where any decrease in the capillary pressure resulted in a reduction of the oil saturation. Moreover, as the capillary pressure approached zero, the water saturation was relatively high ($S_w = 0.736$), which matches the expected physical behavior as, for the imbibition of a strongly wetting phase, the capillary pressure generally does not reach zero until the wetting-phase saturation is significant [1].

In this case, the spontaneous imbibition was responsible for a major increase in the water saturation. Moreover, as capillary pressure was further decreased and became negative, a residual oil saturation was achieved at $S_{o,r} = 0.095$, which is lower than the weakly water-wet case. That matches the expected physical behavior as compared to the expected residual oil saturation of a strongly water-wet media [1]. Snapshots for this case are shown in Figure 13 (b).

### 4.2.2.2. Weakly oil-wet medium

In this case, the contact angle was set as $\theta \cong 110°$ ($\xi = 0.2$) to represent a weakly oil-wet system. The imbibition simulation results are shown in Figure 12, where the initial reductions of capillary pressure did not increase the water saturation, and there was no spontaneous imbibition. In this case, the water phase is the non-wetting phase. Hence it was harder for water to start invading the domain. As forced imbibition started, the water saturation began to increase gradually until most of the domain was filled with water. The residual oil saturation $S_{o,r} = 0.293$ for this case is a realistic value [1].

Moreover, we noted that the residual oil saturation value is larger than the previous two cases of weakly and strongly water wetting medium, which matches the expected physical behavior as oil has a higher tendency to stick to solid walls. Hence, larger volumes of oil were trapped in the domain during the imbibition process. Snapshots of this imbibition process are shown in Figure 13 (c), where a gradual increase in the water saturation could be observed until reaching $S_{o,r}$. Such a displacement process is encountered during production from oil reservoirs using water flooding, where water would trap oil in the reservoir and block its flow to production wells.

## 5. SUMMARY

In this work, we introduced an improved version of the multicomponent pseudopotential lattice Boltzmann method. The model combines recent enhancements from the literature, namely a modified version of the explicit forcing scheme, special treatments for interaction forces at the domain boundaries, and an improved virtual solid density scheme for the wetting boundary conditions. We also introduced a new inlet and outlet pressure boundary condition. We showed that the presented scheme alleviated many drawbacks of the pseudopotential model. For instance, it achieved a higher fluids viscosity ratio range and allowed independent surface tension tuning. The model suppressed a non-physical behavior where there used to be significant changes in the volume of trapped fluids.

We used the developed model to simulate two main fluid displacement processes in porous media: primary drainage and imbibition. The simulated system comprised of a heterogeneous porous medium where oil and water displaced each other. The results showed that the developed model correctly captured the underlying physical behavior of the system when compared with experimental results. Also, we highlighted that the values of critical parameters such as irreducible water saturation and residual oil saturation were comparable to realistic values. For instance, irreducible water saturation $S_{w,i}$ was 0.2 for a primary drainage process in a weakly water-wet medium, while it was 0.3 for the same process in a strongly water-wet medium. Moreover, critical oil saturation $S_{o,r}$ was 0.145 for an imbibition process in a weakly water-wet medium, while it was 0.293 in a weakly oil-wet medium.

We observed that the model captured some unique physical phenomena, such as the capillary pressure bump, which occurs in petroleum reservoirs due to the heterogeneity of the porous medium. Furthermore, we altered the wettability of the medium and investigated its effect on the capillary pressure-saturation relationships for primary drainage and imbibition.

Several points could be addressed in future work to explore the performance of the developed model. For instance, the model could be implemented to simulate three-dimensional (3D) geometries to assess its performance. Moreover, the fluid displacement process could be simulated in realistic geometries of porous media. Such geometries could be obtained using micro-computed tomography (μCT) scans [25]. The capillary pressure-saturation relationships could be directly compared with experimental results from SCAL. For the latter, the conversion would be needed between lattice units and physical units, in addition to considerable computational resources to accommodate the simulated domain size.

## CONFLICTS OF INTEREST

The authors have no conflicts to disclose.

## ACKNOWLEDGMENTS

M. Sedahmed and H. A. Warda thanks the High-Performance Computing (HPC) team from Bibliotheca Alexandria (BA) for their support in developing the computer codes used in the present simulations. R. C. V. Coelho acknowledges financial support from the Portuguese

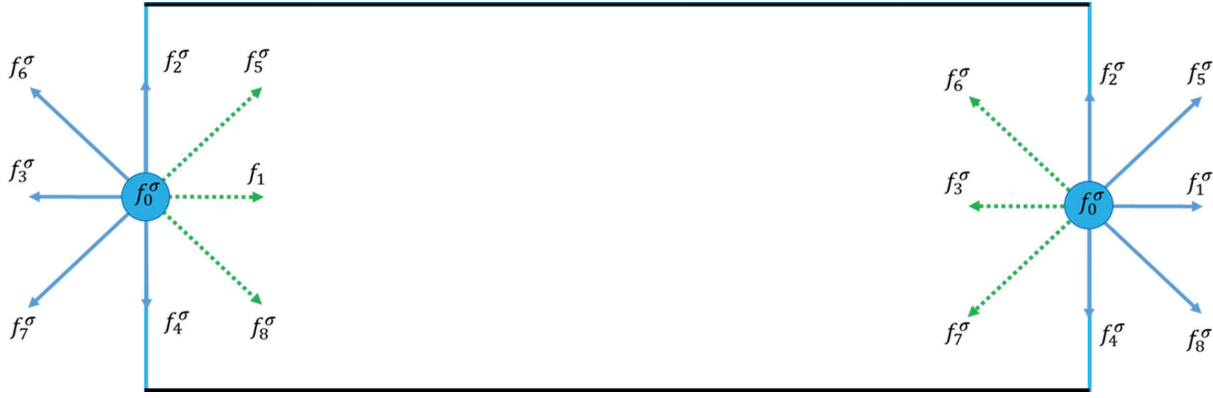

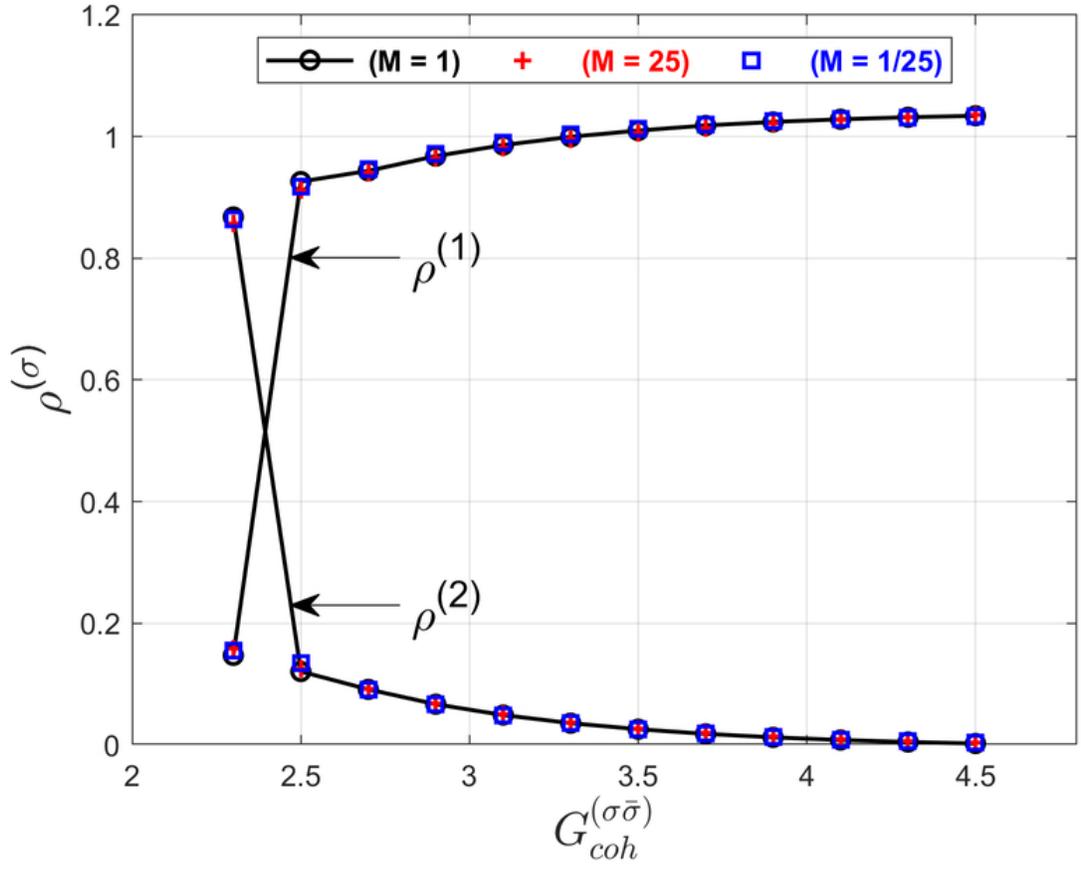

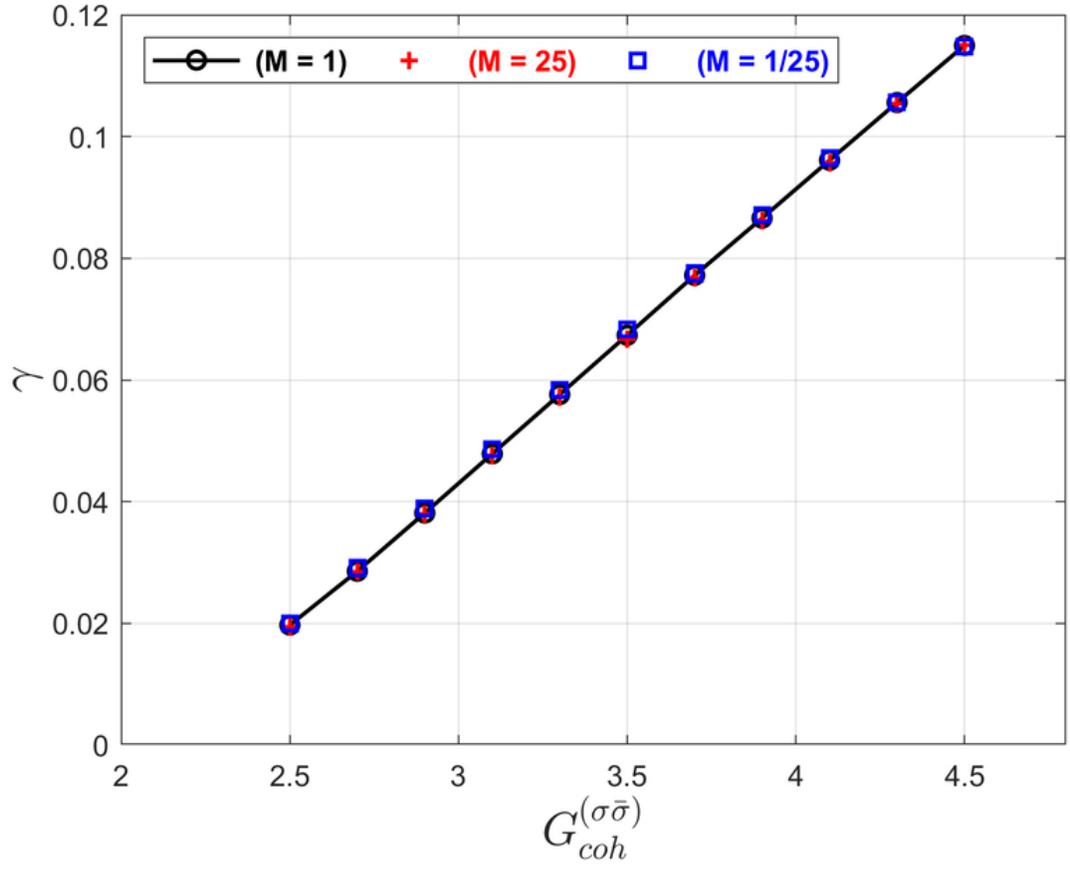

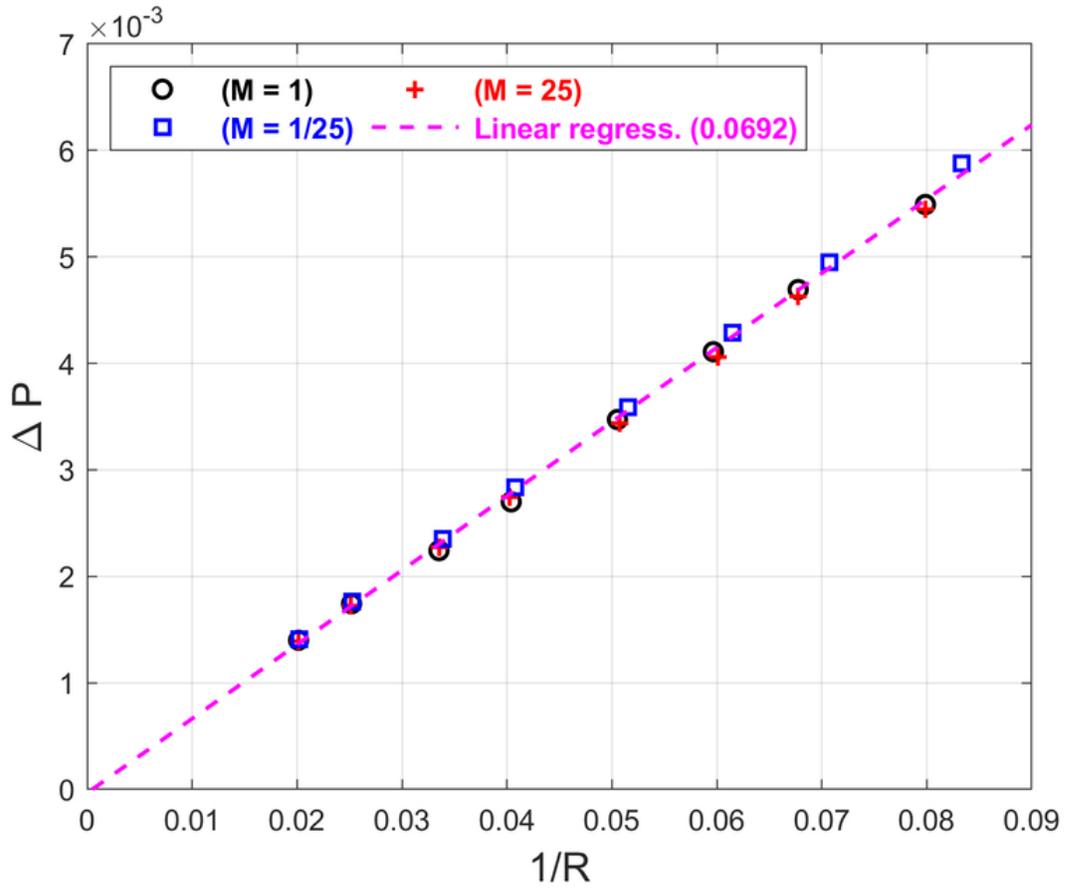

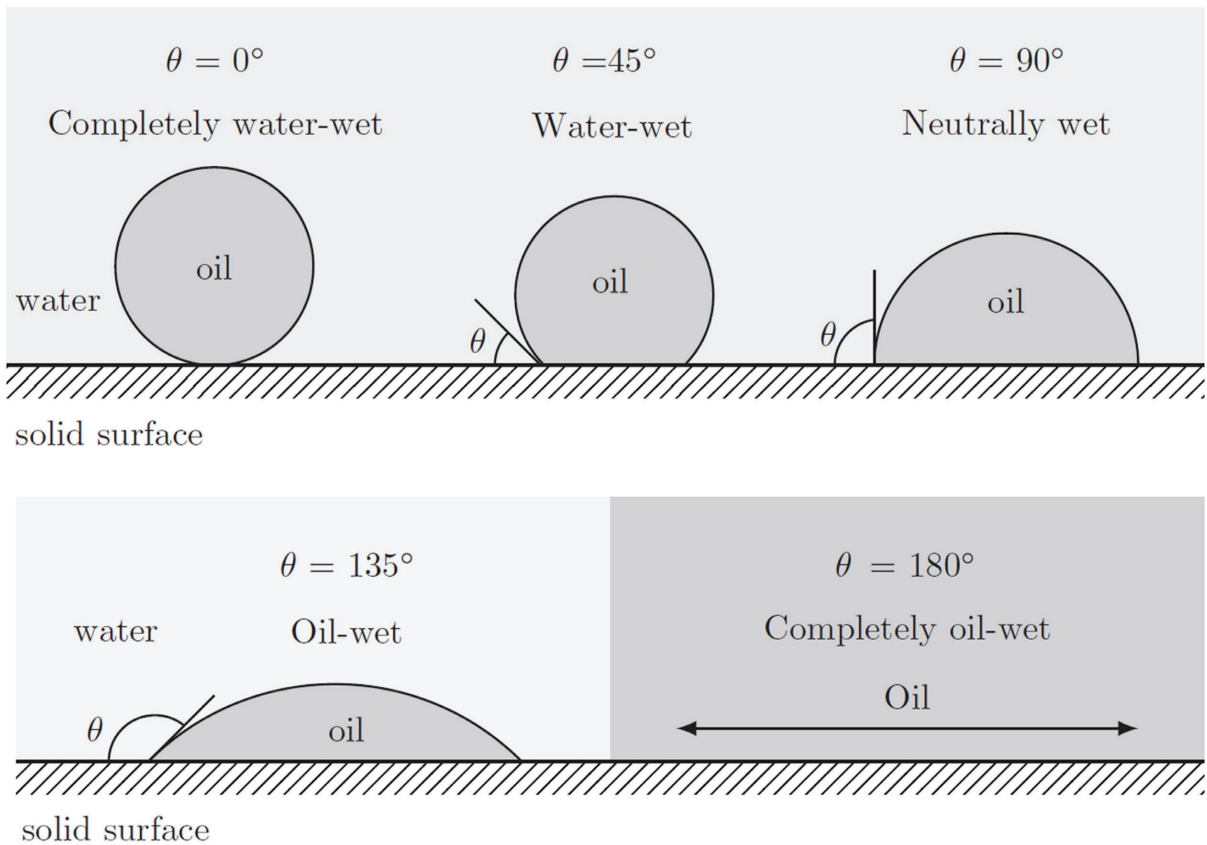

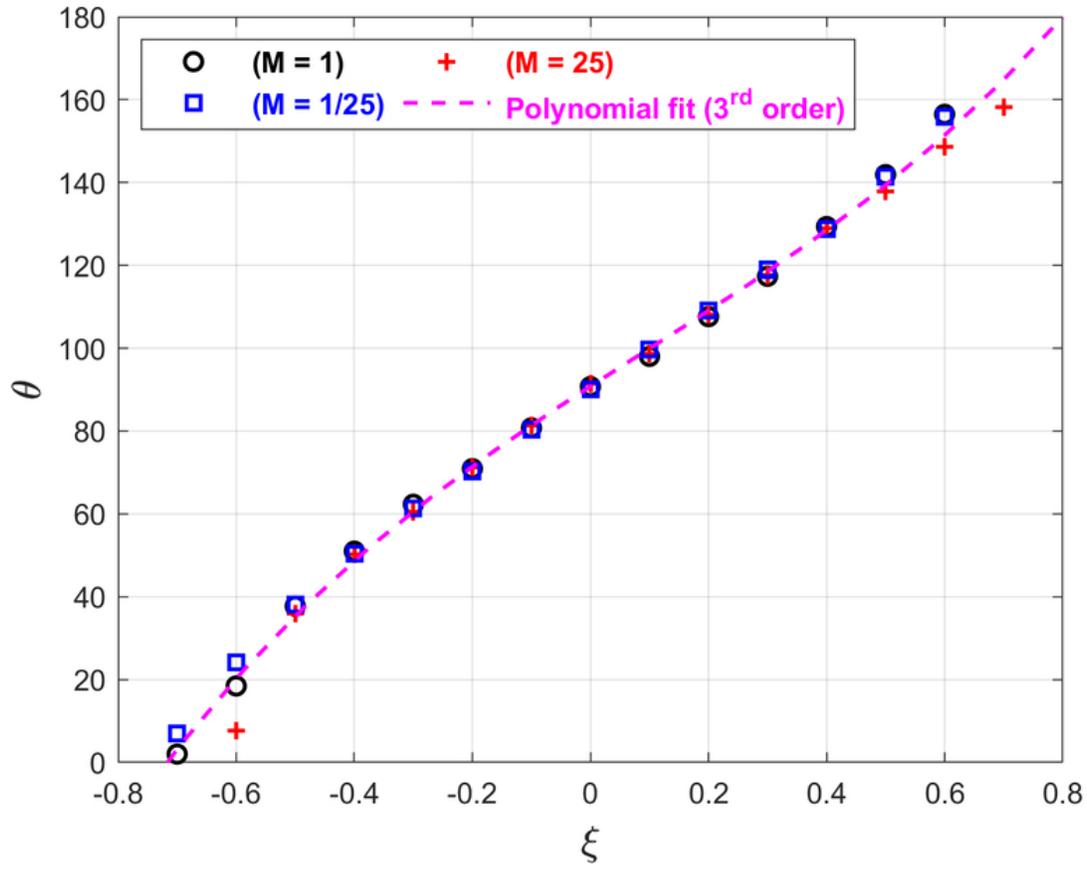

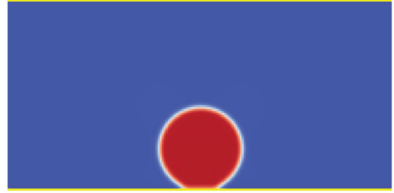

$\xi = -0.6$
$\theta = 7.7°$

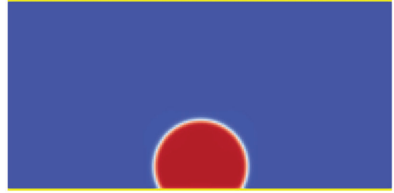

$\xi = -0.3$
$\theta = 60.5°$

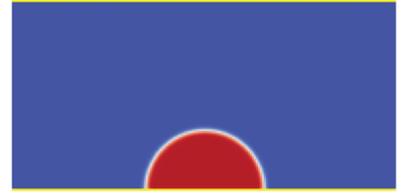

$\xi = 0.0$
$\theta = 91.5°$

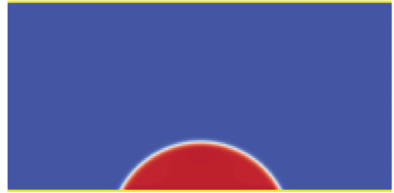

$\xi = 0.3$
$\theta = 117.6°$

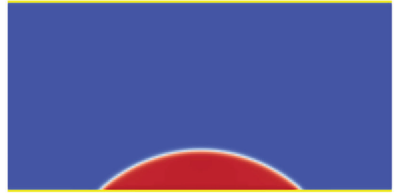

$\xi = 0.5$
$\theta = 137.8$

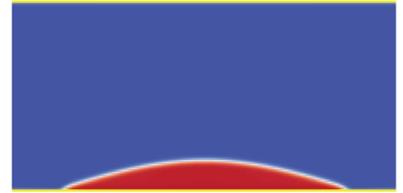

$\xi = 0.7$
$\theta = 158.3$

Physics of Fluids

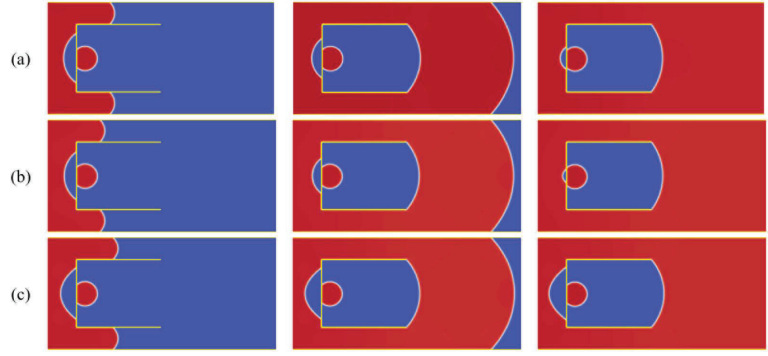

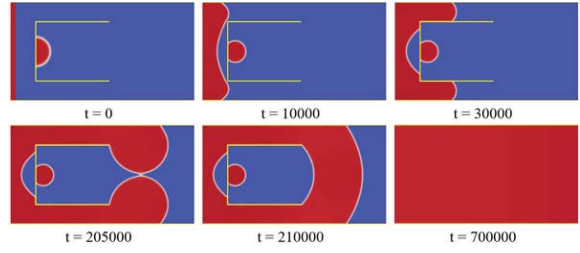

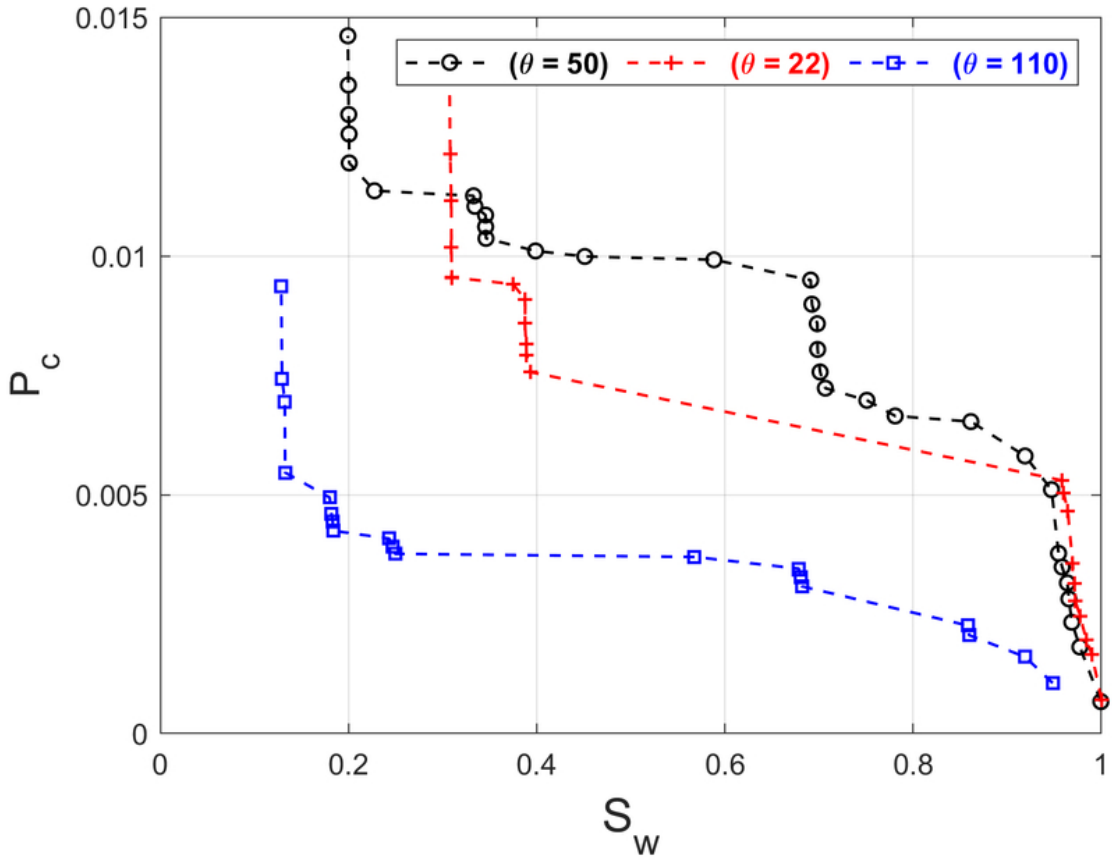

Physics of Fluids

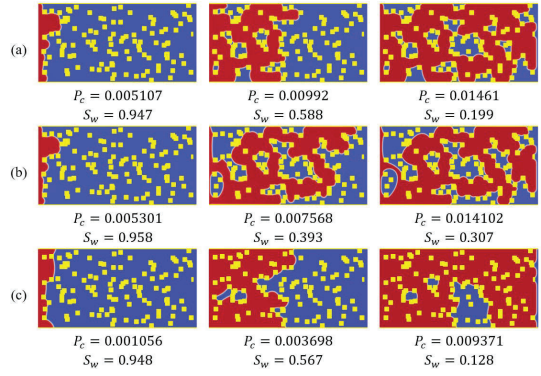

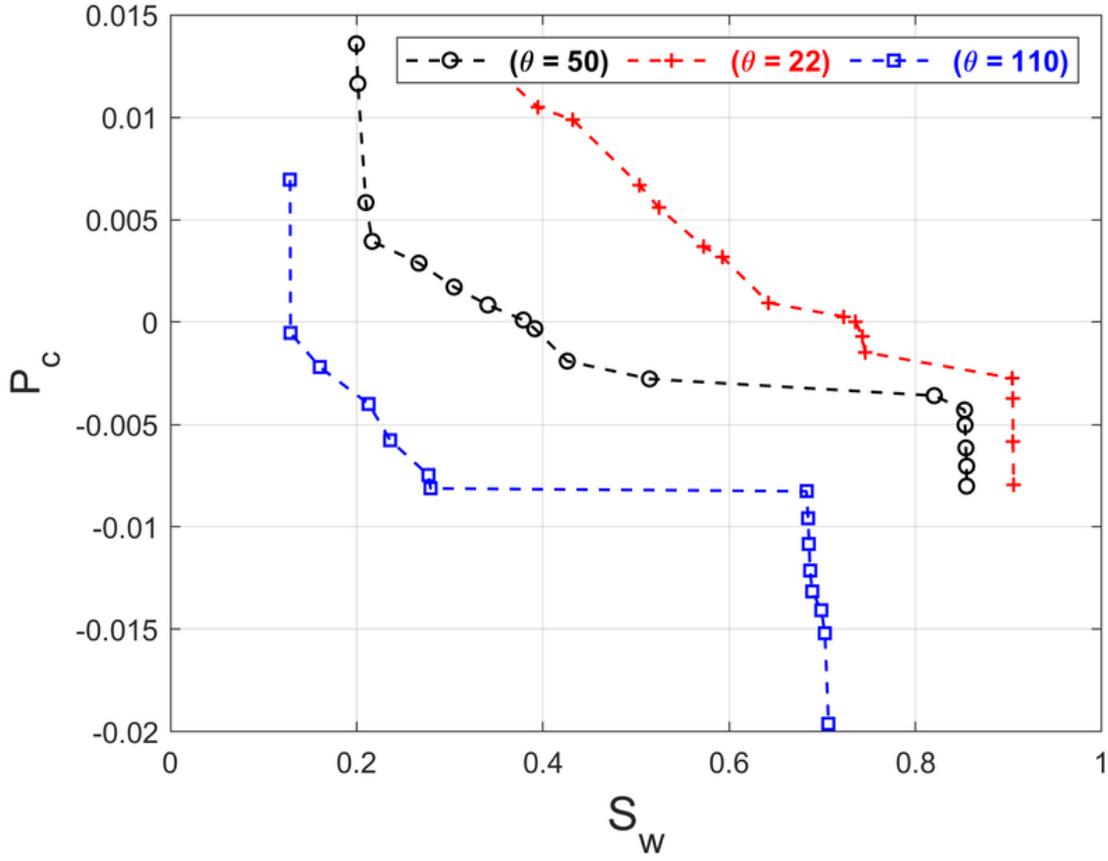

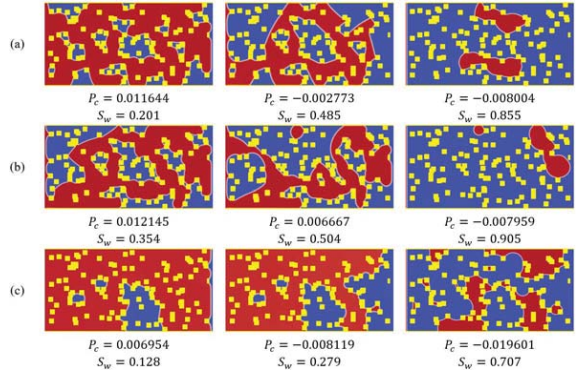

(a)

$P_c = 0.011644$
$S_w = 0.201$

$P_c = -0.002773$
$S_w = 0.485$

$P_c = -0.008004$
$S_w = 0.855$

(b)

$P_c = 0.012145$
$S_w = 0.354$

$P_c = 0.006667$
$S_w = 0.504$

$P_c = -0.007959$
$S_w = 0.905$

(c)

$P_c = 0.006954$
$S_w = 0.128$

$P_c = -0.008119$
$S_w = 0.279$

$P_c = -0.019601$
$S_w = 0.707$